\documentclass[12pt]{article}
\usepackage{graphicx}
\usepackage{amsmath}
\usepackage{amssymb}
\usepackage{caption2}
\begin{document}
\bibliographystyle{prsty}
\begin{center}
{\large {\bf \sc{  Reanalysis of  the $(0^+,1^+)$   states  $B_{s0}$ and $B_{s1}$ with QCD sum rules }}} \\[2mm]
Zhi-Gang Wang \footnote{E-mail,wangzgyiti@yahoo.com.cn.  }     \\
 Department of Physics, North China Electric Power University,
Baoding 071003, P. R. China

\end{center}

\begin{abstract}
In this article, we calculate the masses and decay constants
  of the $P$-wave strange-bottomed  mesons $B_{s0}$ and $B_{s1}$
   with the QCD sum rules,  and observe that the central values of the masses
   of the $B_{s0}$ and $B_{s1}$ are
  smaller than the corresponding
$BK$ and $B^*K$ thresholds respectively,  the strong decays
$B_{s0}\rightarrow BK$ and $B_{s1}\rightarrow B^*K$ are
kinematically  forbidden. They can decay through the isospin
violation precesses $B_{s0}\rightarrow B_s\eta\rightarrow B_s\pi^0$
and $B_{s1}\rightarrow B_s^*\eta\rightarrow
 B_s^*\pi^0$. The bottomed   mesons $B_{s0}$ and $B_{s1}$, just like their charmed
cousins $D_{s0}(2317)$ and $D_{s1}(2460)$, maybe very narrow.
\end{abstract}



\section{Introduction}

According to the heavy quark effective theory \cite{Neubert94}, in
the limit  $m_Q \rightarrow \infty$, the heavy quark decouples from
the light degrees of freedom. $\vec{J}=\vec{S}_Q+\vec{j_q}$,
$\vec{j_q}=\vec{S_q}+\vec{L}$, where $\vec{S}_Q$ and $\vec{S}_q$ are
the spins of the heavy and light quarks respectively,
 $\vec{L}$ is the total angular momentum. For the
$P$-wave strange-bottomed  states, there are two degenerate
doublets: the $j_q=\frac{1}{2}$
  states $B_{s0}$ and $B_{s1}$, and the $j_q=\frac{3}{2}$ states $B^*_{s1}$ and $B_{s2}^*$.
   If kinematically
allowed, the states with $j_q=\frac{1}{2}$ can decay via an $S$-wave
transition, while the $j_q=\frac{3}{2}$ states undergo a $D$-wave
transition; the decay widths of the states with $j_q=\frac{1}{2}$
are expected to be much broader than the corresponding
$j_q=\frac{3}{2}$ states.

 Recently, the CDF Collaboration  reports the first observation of two narrow resonances
consistent with the orbitally excited $P$-wave $B_s$ mesons using
$1$ $\mathrm{fb^{-1}}$ of $p\overline{p}$ collisions at $\sqrt{s} =
1.96 \,\rm{TeV} $ collected with the CDF II detector at the Fermilab
Tevatron \cite{CDF}. The masses  are $M(B^*_{s1})=( 5829.4 \pm 0.7)
\,\rm{MeV}$ and $M(B_{s2}^*) = (5839.7 \pm 0.7)\,\rm{MeV}$.  The
D0 Collaboration  reports the direct observation of the excited
$P$-wave state $B_{s2}^*$
 in fully reconstructed decays to $B^+K^-$, the mass  is
 $(5839.6 \pm 1.1  \pm 0.7)\, \rm{MeV}$ \cite{D0}. The  $P$-wave $B_s$ states with spin-parity
 $J^P=(0^+,1^+)$ are still lack experimental evidence, they maybe observed at the Tevatron or LHCb.

 \begin{table}
\begin{center}
\begin{tabular}{|c|c|}
\hline\hline Mesons & References   \\ \hline $D,D_s,B,B_s$ &
\cite{RRY1980-BDs,RRY1981-T,Aliev1983-DB,Dominguez1987-DB,Narison1987-DB,Reinders1988-B,
Narison1988-DBspva,Narison1993-B,Narison2001-DB,Penin2002-DB,Jamin2002-BBs,
Hayashigaki04-T, Narison2005-DBx}
\\\hline
$D^*,D_s^*,B^*,B_s^*$ & \cite{RRY1981-V,RRY1981-T,Narison1988-DBspva,Hayashigaki04-T,Narison2005-DBx}\\
\hline $D_0,D_1,D_{s0},D_{s1}$ &
\cite{Hayashigaki04-T,Narison2005-DBx}\\
\hline
$B_0,B_1$($B_{s0},B_{s1}$) & \cite{RRY1981-T,Narison1988-DBspva,Narison2005-DBx}(\cite{RRY1981-T})\\
\hline
 \hline
\end{tabular}
\end{center}
\caption{ The works on the charmed and bottomed heavy-light mesons
with the QCD sum rules. }
\end{table}

The masses of the  $B_s$ states with $J^P=(0^+,1^+)$ have been
estimated with the potential models, heavy quark effective theory
and lattice QCD
 \cite{Ebert98,Godfrey91,Bardeen03,Colangelo07,Green04,Eichten01,Vijande07,Lee1,Lee2},
 they are listed in Table.2, from the table, we can see that  the values are different from each other.

There have been many works on the  charmed and bottomed heavy-light
mesons with the QCD sum rules, they are classified roughly  in
Table.1. From the table, we can see that most of the works focus on
the pseudoscalar mesons, the works on the
 scalar and axial-vector mesons are few.

In Ref.\cite{RRY1981-T}, Reinders, Yazaki and Rubinstein study the
strange-bottomed $(0^+,1^+)$ states  $(B_{s0},B_{s1})$ with the
moment  sum rules, due to the large threshold parameters $s_0=70\,
\rm{GeV}^2$, they obtain large values $M_{S}=6.29\, \rm{GeV}$ and
$M_{A}=6.34\,\rm{GeV}$, which are larger than the experimental
values for the corresponding $(1^+,2^+)$ states about
$0.5\,\rm{GeV}$ \cite{CDF,D0}.

 Although  we  can obtain useful predictions for the masses of
 the  strange-bottomed $(0^+,1^+)$ states  $(B_{s0},B_{s1})$ from
  the works of Narison  with the symmetry properties \cite{Narison1988-DBspva,Narison2005-DBx},
   it is valuable to study the strange-bottomed $(0^+,1^+)$ states   with the QCD sum rules directly.

In this article, we  calculate the masses and decay constants of the
 $B_{s0}$ and $B_{s1}$
 with the QCD sum rules
\cite{SVZ79,Reinders85}. In the QCD sum rules, the operator product
expansion is used to expand the time-ordered currents into a series
of quark and gluon condensates  which parameterize the long distance
properties of  the QCD vacuum. Based on current-hadron duality, we
can obtain copious information about the hadronic parameters at the
phenomenological side.

The article is arranged as follows:  we derive the QCD sum rules for
the masses and decay constants of  the $B_{s0}$ and $B_{s1}$ in
section 2; in section 3, numerical results and discussions; section
4 is reserved for conclusion.

\section{QCD sum rules for  the $B_{s0}$ and $B_{s1}$}
In the following, we write down  the two-point correlation functions
$\Pi_{\mu\nu}(p)$ and $\Pi_{SS}(p)$ in the QCD sum rules,
\begin{eqnarray}
\Pi_{\mu\nu}(p)&=&i\int d^4x e^{ip \cdot x} \langle
0|T\left\{J^A_\mu(x)J^{A\dagger}_\nu(0)\right\}|0\rangle \, ,  \\
\Pi_{SS}(p)&=&i\int d^4x e^{ip \cdot x} \langle
0|T\left\{J_S(x)J^{\dagger}_S(0)\right\}|0\rangle \, ,  \\
J^A_\mu(x)&=&\bar{s}(x) \gamma_\mu \gamma_5 b(x) \, ,\nonumber\\
J_S(x)&=&\bar{s}(x)  b(x) \, .\nonumber
\end{eqnarray}
The correlation function  $\Pi_{\mu\nu}(p)$ can be decomposed as
follows,
\begin{eqnarray}
\Pi_{\mu\nu}(p)&=&(-g_{\mu\nu}+\frac{p_\mu
p_\nu}{p^2})\Pi_1(p^2)+p_\mu p_\nu \Pi_0(p^2) +\cdots
\end{eqnarray}
due to  Lorentz covariance, we choose the tensor  structure
$-g_{\mu\nu}+\frac{p_\mu p_\nu}{p^2}$ for analysis.

After performing the standard procedure  of the QCD sum rules, we
obtain the following two sum rules,
\begin{eqnarray}
f_{S}^2 M_{S}^2 e^{-\frac{M_{S}^2}{M^2}}
&=&\frac{3}{8\pi^2}\int_{m_b^2}^{s_S^0} ds
e^{-\frac{s}{M^2}}s\left(1-\frac{m_b^2}{s}\right)^2
 \left[1- \frac{2 m_b m_s}{s-m_b^2}
+\frac{4}{3}\frac{\alpha_s(s)}{\pi}R_0(\frac{m_b^2}{s})\right]
\nonumber\\
&&+ e^{-\frac{m_b^2}{M^2}} \left[ m_b\langle\bar{s}s\rangle
+\frac{m_s\langle\bar{s}s\rangle}{2}\left(1+\frac{m_b^2}{M^2}\right)
 + \frac{1}{12} \langle\frac{\alpha_sGG}{\pi}\rangle \right.
\nonumber\\
&&\left.+  \frac{m_b\langle\bar{s}g_s\sigma  G
s\rangle}{2M^2}\left(1- \frac{m_b^2}{2M^2}\right) -
\frac{16\pi\alpha_s\langle\bar{s}s\rangle^2}{27M^2} \left(1-
\frac{m_b^2}{4M^2}- \frac{m_b^4}{12M^4}\right) \right], \nonumber\\
\end{eqnarray}

 \begin{eqnarray}
 f_{A}^2
M_{A}^2 e^{-\frac{M_{A}^2}{M^2}}
&=&\frac{1}{8\pi^2}\int_{m_b^2}^{s^{0}_A} ds
e^{-\frac{s}{M^2}}s\left(1-\frac{m_b^2}{s}\right)^2
\left(2+\frac{m_b^2}{s}\right)
 \left[1- \frac{3 m_b m_s s}{(2s+m_b^2)(s-m_b^2)}
 \right.\nonumber\\
 &&\left.+\frac{4}{3}\frac{\alpha_s(s)}{\pi}R_1(\frac{m_b^2}{s})\right]
+ e^{-\frac{m_b^2}{M^2}} \left[  m_b\langle\bar{s}s\rangle +
\frac{m_b^2m_s\langle\bar{s}s\rangle}{2M^2} -\frac{1}{12}
\langle\frac{\alpha_sGG}{\pi} \rangle\right.
\nonumber\\
&&\left.  - \frac{m_b^3\langle\bar{s}g_s\sigma G s\rangle }{4M^4}
+\frac{32\pi\alpha_s\langle\bar{s}s\rangle^2}{81M^2}
\left(1+\frac{m_b^2}{M^2}-\frac{m_b^4}{8M^4}\right) \right]\, ,
\end{eqnarray}
where
 \begin{eqnarray}
R_0(x)&=&\frac{9}{4}+2{\rm Li}_2(x)+{\rm ln}x\,{\rm ln}(1-x)
-\frac{3}{2}\,{\rm ln}\frac{1-x}{x}-{\rm ln}(1-x)+x\,{\rm
ln}\frac{1-x}{x}
\nonumber\\
&&-\frac{x}{1-x}{\rm ln}x \, ,\\
R_1(x)&=&\frac{13}{4}+2{\rm Li}_2(x)+{\rm ln}x\,{\rm ln}(1-x)
-\frac{3}{2}\,{\rm ln}\frac{1-x}{x}-{\rm ln}(1-x)+x\,{\rm
ln}\frac{1-x}{x}
\nonumber\\
&&-\frac{x}{1-x}{\rm ln}x+\frac{(3+x)(1-x)}{2+x}\,{\rm
ln}\frac{1-x}{x} -\frac{2x}{(2+x)(1-x)^2}\,{\rm ln}x
\nonumber\\
&&-\frac{5}{2+x}-\frac{2x}{2+x}-\frac{2x}{(2+x)(1-x)} \, ,
 \end{eqnarray}
  ${\rm Li}_2(x)=-\int_0^x dt\,t^{-1}{\rm ln}(1-t)$,
 $GG=G_{\mu\nu}G^{\mu\nu}$, $\sigma
 G=\sigma_{\mu\nu}G^{\mu\nu}$, $\frac{\alpha_s(s)}{4\pi}=\frac{1}{\frac{23}{3}\rm{log}(s/\Lambda_{QCD}^2)}$,
  $\Lambda_{QCD}=226\rm{MeV}$ and $\alpha_s(1\rm{GeV})=0.514$
  \cite{Buras98}. We have used the standard definitions for the decay constants $f_S$ and
  $f_A$,
\begin{eqnarray}
f_{S}M_{S}  &=& \langle 0|J_S(0)|B_{s0}(p)\rangle\, ,\nonumber \\
f_{A}M_{A}\epsilon_\mu  &=& \langle 0|J^A_\mu(0)|B_{s1}(p)\rangle\,
.
\end{eqnarray}
  The spectral densities of the heavy-light $(0^+,1^+)$ mesons
  can be obtained from the corresponding  $(0^-,1^-)$
  mesons with the simple replacement $m_Q\rightarrow -m_Q$, where
  the $m_Q$ stands for the masses of the heavy quarks. Although the expressions
  from different references  have minor differences  from each other,
  the contributions from those terms are usually of minor importance.  We have consulted  the analytical expressions
  presented in Refs.\cite{RRY1981-V,RRY1980-BDs,RRY1981-T,Aliev1983-DB,Narison2005-DBx,Jamin93-CF,Generalis1990-CF,Pfannmoller-CF}
  to obtain our main results.

 Differentiating  the Eqs.(4-5) with respect to  $\frac{1}{M^2}$, then eliminate the
 quantities $f_{S}$ and $f_{A}$, we can obtain two sum rules for
 the masses of the $B_{s0}$ and $B_{s1}$, respectively.

\section{Numerical results and discussions}
The input parameters are taken to be the standard values $\langle
\bar{q}q \rangle=-(0.24\pm 0.01\, \rm{GeV})^3$, $\langle \bar{s}s
\rangle=(0.8\pm 0.2 )\langle \bar{q}q \rangle$, $\langle
\bar{s}g_s\sigma Gs \rangle=m_0^2\langle \bar{s}s \rangle$,
$m_0^2=(0.8 \pm 0.2)\,\rm{GeV}^2$, $\langle \frac{\alpha_s
GG}{\pi}\rangle=(0.33\,\rm{GeV})^4 $, $m_s=(0.14\pm0.01)\,\rm{GeV}$
 and $m_b=(4.7\pm0.1)\,\rm{GeV}$ at the energy scale about $\mu=1\, \rm{GeV}$
\cite{SVZ79,Reinders85,Ioffe2005}.

In Refs.\cite{Hayashigaki04-T,Narison2005-DBx}, different
predictions for the masses of the scalar mesons $D_{s0}$ and $D_{0}$
are obtained due to different values of the threshold parameters
$s_0$ and charmed quark mass $m_c$. The mass $m_b$ is very large
comparing with the $m_c$, the uncertainty about $0.1 \, \rm{GeV}$
cannot change the predictions remarkably. We choose the threshold
parameters $s_0$ with the guide of the experimental data and
predictions of the potential models to reduce the uncertainty.

In 2006, the BaBar Collaboration observed a new $c\bar{s}$ state
$D_s(2860)$ with the mass $M=(2856.6 \pm 1.5 \pm 5.0)\,\rm{MeV}$,
width  $\Gamma=(48 \pm 7 \pm 10)\,\rm{MeV}$ and possible spin-parity
$0^+$, $1^-$, $2^+, \cdots$ \cite{BaBar0607}. It has been
interpreted as the first radial excitation of the $0^+$ state
$D_{s0}(2317)$ in Refs.\cite{2Ds-1,2Ds-2}, although other
identifications  are not excluded. The energy gap between the $2P$
and $1P$ scalar $c\bar{s}$ states is about $\delta
M_S=0.539\,\rm{GeV}$. In 2007, the Belle Collaboration observed  a
new resonance $D_{s}(2700)$
 in the decay $B^{+} \to \bar{D}^{0} D_{s}(2700) \to \bar{D}^{0}
D^{0} K^{+}$,  which has  the mass  $M_V=2708 \pm 9 ^{+11}_{-10}
\,\rm{MeV}$, width $\Gamma_V = 108 \pm 23 ^{+36}_{-31} \,\rm{MeV}$,
and spin-parity $1^{-}$\cite{Belle0707}. They interpret the
 $D_s (2700)$ as a $c\bar{s}$ meson, the potential models
 predict a  radially excited $2^3S_1$ ($c\bar{s}$) state
with a mass about $(2710-2720)\,\rm{MeV}$ \cite{2Dv-1,2Dv-2}, so the
energy gap between the $2S$ and $1S$ vector $c\bar{s}$ states is
about $\delta M_V=0.596\,\rm{GeV}$.

If the masses of the $P$-wave strange-bottomed mesons are of the
same order (about $5.8\,\rm{GeV}$ \cite{CDF,D0}) and the energy gap
between the ground state and the first radially  excited state is
about $0.5\,\rm{GeV}$ (just like the $c\bar{s}$ mesons), we can make
a rough estimation for the masses of the first radially  excited
$(0^+,1^+)$ strange-bottomed states, $M_r\approx
(5.8+0.5)\,\rm{GeV}$. The threshold parameters should be chosen as
$s_0< M_r^2\approx40\,\rm{GeV}^2$, which are consistent with the
predictions of potential models \cite{Eichten01}.

The threshold parameters can be taken as
$s_S^0=(37\pm1)\,\rm{GeV}^2$ and $s^0_A=(38\pm1)\,\rm{GeV}^2$, which
are   below the corresponding masses of the first radially  excited
states, $M_{Sr}=6.264\,\rm{GeV}$ for the $B_{s0}$ and
$M_{Ar}=6.296\,\rm{GeV}$ for the $B_{s1} $  in the potential model,
see Ref.\cite{Eichten01}. The threshold parameters $s_0=70\,
\rm{GeV}^2$ chosen by Reinders, Yazaki and Rubinstein in
Ref.\cite{RRY1981-T} are too large  to make reliable predictions.

\begin{figure}
 \centering
 \includegraphics[totalheight=6cm,width=7cm]{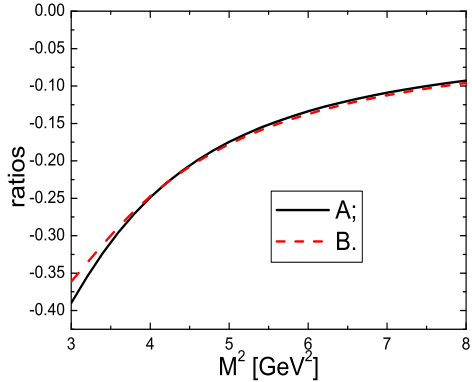}
  \caption{The ratios between the contributions of the non-perturbative terms and the perturbative terms with
  variation of
  the Borel parameter $M^2$,   $A$ for the $B_{s0}$  and $B$ for the $B_{s1}$. }
\end{figure}

The Borel parameters are taken as
 $M^2=(4-6)\,\rm{GeV}^2$  for the $B_{s0}$ (see Eq.(4)) and $M^2=(5-7)\,\rm{GeV}^2$ for the $B_{s1}$ (see Eq.(5)).
  In those regions, the
contributions from the pole terms are larger than $50\%$;
furthermore, the dominating contributions come from the perturbative
terms. In Fig.1, we plot the ratios  between the contributions of
the non-perturbative terms (come from the vacuum condensate $\langle
\bar{s}s\rangle$ mainly) and the perturbative terms with variation
of the Borel parameter $M^2$. In the region $M^2=(4-7) \,
\rm{GeV}^2$, the ratios are about $-(0.10-0.25)$. The  criterions
(pole dominance and convergence of the operator product expansion)
of the QCD sum rules are well satisfied, our predictions are robust.

\begin{figure}
 \centering
 \includegraphics[totalheight=6cm,width=7cm]{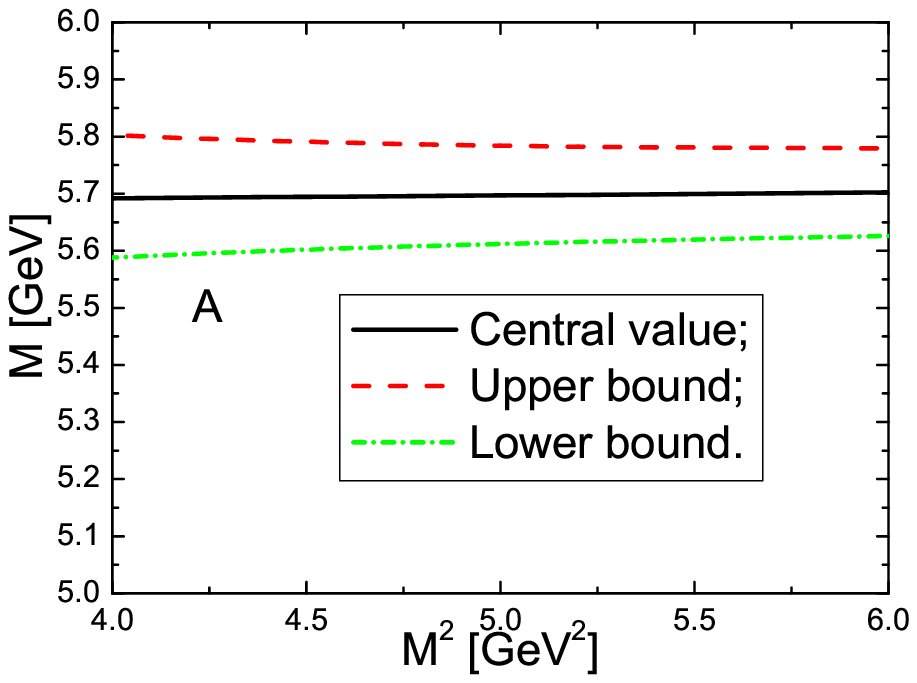}
 \includegraphics[totalheight=6cm,width=7cm]{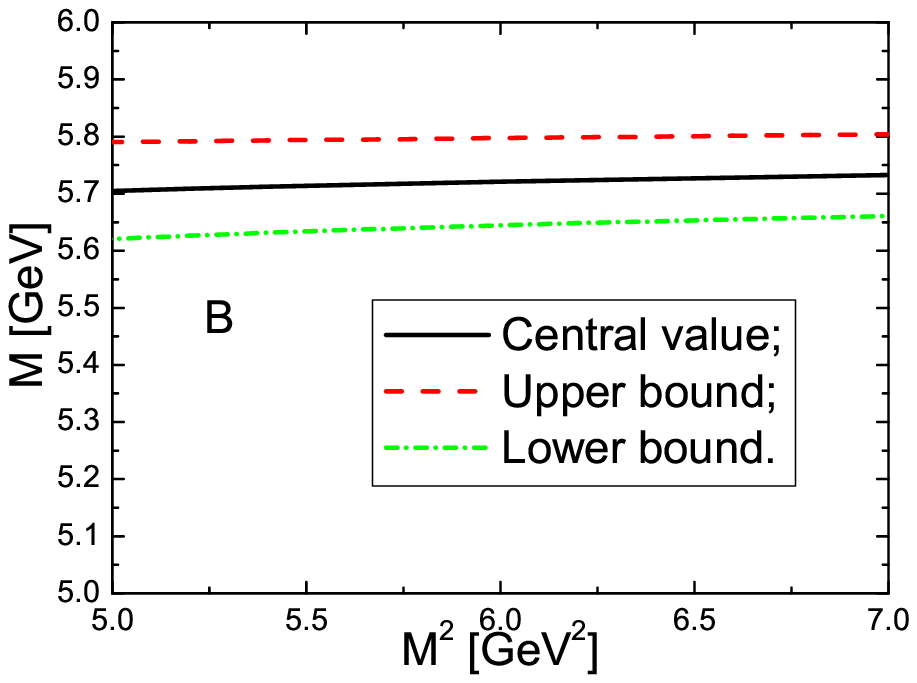}
 \caption{$M_{S}$($A$) and $M_{A}$($B$) with variation of the Borel parameter $M^2$. }
\end{figure}

\begin{figure}
 \centering
 \includegraphics[totalheight=6cm,width=7cm]{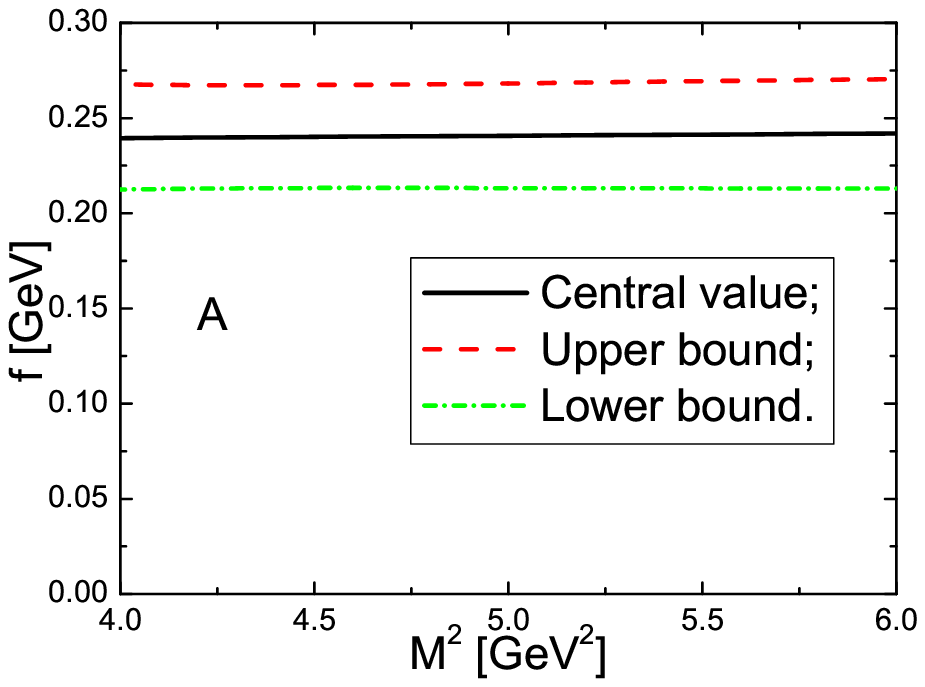}
 \includegraphics[totalheight=6cm,width=7cm]{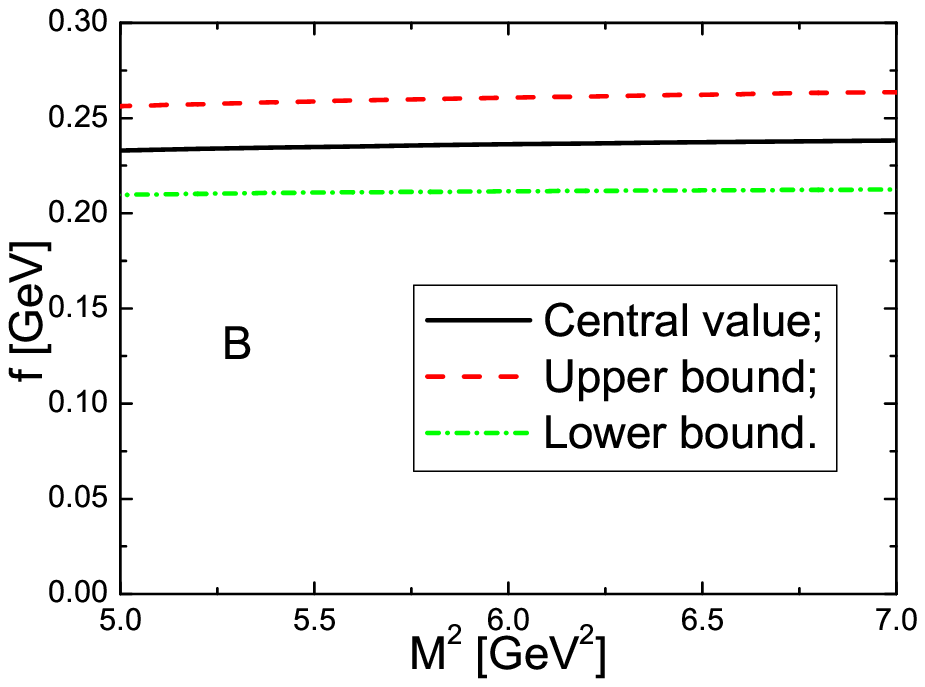}
 \caption{$f_{S}$($A$) and $f_{A}$($B$) with variation of the Borel parameter $M^2$. }
\end{figure}

\begin{table}
\begin{center}
\begin{tabular}{|c|c|c|}
\hline\hline & $M_{S}(\rm{GeV})$& $M_{A}(\rm{GeV})$\\ \hline
      \cite{Ebert98}  &$5.841$ &$5.831$\\ \hline
      \cite{Godfrey91}& $5.830$ &$5.786$\\      \hline
     \cite{Bardeen03}& $5.718$ &$5.765$\\     \hline
    \cite{Colangelo07} &  $5.71$ &$5.77$\\ \hline
     \cite{Green04}&  $5.756\pm0.031$ &$5.804\pm0.031$\\ \hline
     \cite{Eichten01} &  $5.804$ &$5.842$\\ \hline
      \cite{Vijande07} &  $5.679$ &$5.713$\\ \hline
       \cite{Nowak04} &  $ 5.689 $ &$5.734$\\ \hline
   This work & $5.70\pm0.11$ &$5.72\pm0.09$\\ \hline
       \hline
\end{tabular}
\end{center}
\caption{ Theoretical estimations for the masses of the $B_{s0}$ and
$B_{s1}$ from different models. }
\end{table}
Taking into account all uncertainties of the input parameters,
finally we obtain the values of the masses and decay constants of
 the $P$-wave heavy mesons $B_{s0}$ and $B_{s1}$, which are
shown in Figs.(2-3), respectively,
\begin{eqnarray}
M_{S}&=&(5.70\pm0.11)\,\rm{GeV} \, , \nonumber\\
M_{A}&=&(5.72\pm0.09)\,\rm{GeV} \, , \nonumber\\
f_{S}&=&(0.24\pm 0.03)\,\rm{GeV} \, , \nonumber\\
f_{A}&=&(0.24\pm 0.02)\,\rm{GeV} \, .
\end{eqnarray}
From the experimental data \cite{PDG}, $m_B=5.279\,\rm{GeV}$,
$m_K=0.495\,\rm{GeV}$ and  $m_{B^*}=5.325\,\rm{GeV}$, we can see
that the central values are below the corresponding  $BK$ and $B^*K$
thresholds respectively, $m_{BK}=5.774\,\rm{GeV}$ and
$m_{B^*K}=5.820\,\rm{GeV} $. The strong decays $B_{s0}\rightarrow
BK$ and $B_{s1}\rightarrow B^*K$ are kinematically  forbidden, the
$P$-wave heavy mesons  $B_{s0}$ and $B_{s1}$ can decay through the
isospin violation precesses $B_{s0}\rightarrow B_s\eta\rightarrow
B_s\pi^0$ and $B_{s1}\rightarrow B_s^*\eta\rightarrow
 B_s^*\pi^0$ respectively. The   $\eta$ and $\pi^0$ transition matrix is very small according to
   Dashen's
 theorem \cite{Dashen},
 \begin{eqnarray}
 t_{\eta\pi}&=&\langle \pi^0 |\mathcal {H}
 |\eta\rangle=-0.003\,\rm{GeV}^2\, ,
 \end{eqnarray}
the $P$-wave bottomed mesons $B_{s0}$ and $B_{s1}$, just like their
charmed cousins $D_{s0}(2317)$ and $D_{s1}(2460)$, maybe very narrow
\cite{Swanson06R,Colangelo04R}.

\section{Conclusion}
In this article, we calculate the masses and decay constants
  of the $P$-wave strange-bottomed  mesons $B_{s0}$ and $B_{s1}$
   with the QCD sum rules,  and observe  that the central values of the masses   are
  smaller than the corresponding
$BK$ and $B^*K$ thresholds respectively,  the strong decays
$B_{s0}\rightarrow BK$ and $B_{s1}\rightarrow B^*K$ are
kinematically  forbidden. They can decay through the isospin
violation precesses $B_{s0}\rightarrow B_s\eta\rightarrow B_s\pi^0$
and $B_{s1}\rightarrow B_s^*\eta\rightarrow
 B_s^*\pi^0$.
The bottomed mesons $B_{s0}$ and $B_{s1}$, just like their charmed
cousins $D_{s0}(2317)$ and $D_{s1}(2460)$, maybe very narrow.

\section*{Acknowledgements}
This  work is supported by National Natural Science Foundation,
Grant Number 10405009, 10775051, and Program for New Century
Excellent Talents in University, Grant Number NCET-07-0282.


\begin{thebibliography}{99}

\bibitem{Neubert94} M. Neubert, Phys. Rept. {\bf 245} (1994) 259.

\bibitem{CDF}  T. Aaltonen et al, Phys. Rev. Lett. {\bf 100} (2008) 082001.

\bibitem{D0} V. Abazov et al, Phys. Rev. Lett. {\bf 100} (2008) 082002.


\bibitem{Ebert98} D. Ebert, V. O. Galkin, and R. N. Faustov, Phys. Rev. {\bf D57} (1998) 5663.



\bibitem{Godfrey91}  S. Godfrey and R. Kokoski, Phys. Rev. {\bf D43} (1991) 1679.

\bibitem{Bardeen03}  W. A. Bardeen, E. J. Eichten and C. T. Hill, Phys. Rev. {\bf D68} (2003)
054024.
\bibitem{Colangelo07} P. Colangelo, F. De Fazio and R. Ferrandes,
Nucl. Phys. Proc. Suppl. {\bf 163} (2007) 177.

 \bibitem{Green04} A. M. Green et al, Phys. Rev. {\bf D69} (2004) 094505.

\bibitem{Eichten01} M. Di Pierro and E. Eichten, Phys. Rev. {\bf D64} (2001)
114004.

\bibitem{Vijande07} J. Vijande, A. Valcarce and F. Fernandez,
Phys. Rev. {\bf D77} (2008) 017501.

\bibitem{Nowak04} M. A. Nowak, M. Rho and I. Zahed, Acta. Phys. Polon. {\bf B35} (2004)
2377.

\bibitem{Lee1} I. W. Lee, T. Lee, D. P. Min and B. Y. Park, Eur. Phys. J. {\bf C49}
(2007) 737.

\bibitem{Lee2}I. W. Lee and T. Lee, Phys. Rev. {\bf D76} (2007) 014017.



\bibitem{RRY1981-V} L. J. Reinders, S. Yazaki and  H. R. Rubinstein, Phys. Lett. {\bf B103} (1981) 63.

\bibitem{RRY1980-BDs} L. J. Reinders, H. R. Rubinstein and S. Yazaki, Phys. Lett. {\bf 97B} (1980)
257.

\bibitem{RRY1981-T} L. J. Reinders, S. Yazaki  and H. R. Rubinstein, Phys. Lett. {\bf B104} (1981) 305.

\bibitem{Aliev1983-DB} T. M. Aliev and V. L. Eletsky, Sov. J. Nucl. Phys. {\bf 38} (1983)
936.

\bibitem{Dominguez1987-DB} C. A. Dominguez and N. Paver, Phys. Lett. {\bf 197B} (1987)
423.

\bibitem{Narison1987-DB} S. Narison, Phys. Lett. {\bf B198} (1987) 104.


\bibitem{Reinders1988-B} L. J. Reinders, Phys. Rev. {\bf D38} (1988) 947.

\bibitem{Narison1988-DBspva} S. Narison, Phys. Lett. {\bf B210} (1988) 238.

\bibitem{Narison1993-B} S. Narison, Phys. Lett. {\bf B308} (1993) 365.

\bibitem{Narison2001-DB} S. Narison, Phys. Lett. {\bf B520} (2001) 115.

\bibitem{Penin2002-DB} A. A. Penin and M. Steinhauser, Phys. Rev. {\bf D65} (2002) 054006.

\bibitem{Jamin2002-BBs} M. Jamin and B. O. Lange, Phys. Rev. {\bf D65} (2002) 056005.

\bibitem{Hayashigaki04-T} A. Hayashigaki and K. Terasaki, hep-ph/0411285.

\bibitem{Narison2005-DBx} S. Narison, Phys. Lett. {\bf B605} (2005) 319.

\bibitem{SVZ79}  M. A. Shifman, A. I. Vainshtein and V. I. Zakharov,
Nucl. Phys. {\bf B147} (1979) 385, 448.

\bibitem{Reinders85} L. J. Reinders, H.
Rubinstein and S. Yazaki, Phys. Rept. {\bf 127} (1985) 1.

\bibitem{Buras98} A. J. Buras, hep-ph/9806471.


\bibitem{Jamin93-CF} M. Jamin and M. Munz, Z. Phys. {\bf C60} (1993) 569.

\bibitem{Generalis1990-CF} S. C. Generalis, J. Phys. {\bf G16} (1990)
785.

\bibitem{Pfannmoller-CF} J. P. Pfannmoller, hep-ph/0608213.

\bibitem{Ioffe2005} B. L. Ioffe, Prog. Part. Nucl. Phys. {\bf 56} (2006)
232; and references  therein.


\bibitem{BaBar0607} B. Aubert et al, Phys. Rev. Lett. {\bf 97} (2006) 222001.

\bibitem{2Ds-1} E. van Beveren and G. Rupp, Phys. Rev. Lett. {\bf 97} (2006) 202001.

\bibitem{2Ds-2} F. E. Close, C. E. Thomas, O. Lakhina and E. S. Swanson, Phys. Lett. {\bf B647 } (2007) 159.

\bibitem{Belle0707}  J. Brodzicka  et al, Phys. Rev. Lett. {\bf 100} (2008) 092001.


\bibitem{2Dv-1}  S. Godfrey and N. Isgur, Phys. Rev. {\bf D32} (1985)
189.

\bibitem{2Dv-2} F. E. Close, C. E. Thomas, O. Lakhina and E. S. Swanson, Phys. Lett.
{\bf B647}  (2007) 159.


\bibitem{PDG} W.-M. Yao et al, J. Phys. {\bf G33} (2006) 1.



\bibitem{Dashen} R. F. Dashen, Phys. Rev. {\bf 183} (1969) 1245.


\bibitem{Swanson06R} E. S. Swanson, Phys. Rept. {\bf 429} (2006) 243;
and references  therein.

\bibitem{Colangelo04R} P. Colangelo, F. De Fazio and R. Ferrandes, Mod. Phys. Lett. {\bf
A19} (2004) 2083; and references  therein.




\end{thebibliography}
\end{document}